\documentclass[a4paper]{article}

\usepackage{INTERSPEECH2021}
  
\title{Detecting Escalation Level from Speech with Transfer Learning and Acoustic-Lexical Information Fusion}
\name{Ziang Zhou$^1$, Yanze Xu$^1$, Ming Li$^{1,2}$\thanks{corresponding author: Ming Li, ming.li369@dukekunshan.edu.cn}}
\address{
  $^1$Data Science Research Center, Duke Kunshan University, Kunshan, China\\
  $^2$School of Computer Science, Wuhan University, Wuhan, China}
\email{ziang.zhou372@dukekunshan.edu.cn, ming.li369@dukekunshan.edu.cn}

\begin{document}

\maketitle
\begin{abstract}
  Textual escalation detection has been widely applied to e-commerce companies’ customer service systems to pre-alert and prevent potential conflicts. Similarly, in public areas such as airports and train stations, where many impersonal conversations frequently take place, acoustic-based escalation detection systems are also useful to enhance passengers’ safety and maintain public order. To this end, we introduce a system based on acoustic-lexical features to detect escalation from speech, Voice Activity Detection (VAD) and label smoothing are adopted to further enhance the performance in our experiments. Considering a small set of training and development data, we also employ transfer learning on several well-known emotional detection datasets, i.e. RAVDESS, CREMA-D, to learn advanced emotional representations that is then applied to the conversational escalation detection task. On the development set, our proposed system achieves 81.5\% unweighted average recall (UAR)  which significantly outperforms the baseline with 72.2\% UAR.
  

\end{abstract}

\section{Introduction}
Escalation level detection system has been applied in wide ranges of applications, including human-computer interaction and computer-based human-to-human conversation~\cite{Schuller21-TI2}. For instance, there are e-commerce companies \cite{letcher2018automatic} that have been equipped with textual conversational escalation detectors. Once an increasing escalation level of the customers is detected, the special agents will take over and settle their dissatisfaction. This can forestall the conflict from worsening and protects the feelings of the employees. In public areas like transportation centers, information desks, where many impersonal interactions take place, it is also important to detect the potential risk of escalations from conversation to guarantee public security. Therefore, audio escalation level analysis is quite useful and important.

In ComParE 2021 Escalation Sub-task(ESS)~\cite{Schuller21-TI2}, we are provided with datasets of conversation audios recorded on the train and at the information desk. About 400 training audios, with an average length of 5 seconds are used for training. 500 audios from a separate dataset are used for testing. Given datasets with such small scales, it would be challenging to learn sufficient representations from the datasets alone. Thus, we are motivated to learn features that connects to escalation level from a wider source. Since emotion is an obvious indicator in potential conversational conflicts, we have good reasons to assume that the ability to learn emotional features from a clip would be helpful to detect escalations from conversations. Through transfer learning, emotional representations and knowledge are learned before the ESS datasets are applied to fine-tune the emotional recognition model. Recent research on small sample set classification tasks also showed promising results on pattern recognition via transfer learning. \cite{ng2015deep}


\section{Related Works}
\subsection{Audio Conflict Escalation Detection}
Several conflict escalation research has been done in recent years, focusing on the count of overlaps and interruptions in speeches. In \cite{grezes2013let}, the number of overlaps is recorded in the hand-labeled dataset and used in conflict prediction. And \cite{caraty2015detecting} uses a support vector machine (SVM) to detect overlap based on acoustic and prosodic features. Research done by Kim et al. in \cite{kim2012automatic} analyzed the level of conversation escalation based on a corpus composed of French TV debates. They proposed an automatic overlap detection approach to extract features, obtained 62.3\% unweighted accuracy on the corpus. Effective as they may seem, these methods are considered impractical in this Escalation Sub-task. First, the length of audio files in \cite{caraty2015detecting} ranges from 3-30 minutes, and the length of conversation audio in \cite{kim2012automatic} is 30 seconds. While in the ESS task, the average length of the corpus is 5 seconds. Most of the time, a clip only contains a single person’s voice, thus a focus on the overlap detection is likely to be unnecessary. Besides, we didn’t spot a significant difference in overlap frequency among different escalation classes, based on conversation script analysis. Second, with a total training corpus duration of fewer than 30 minutes, it is difficult to effectively learn the overlap-count feature. 

\subsection{Transfer Learning}
In \cite{5973853}, supervised transfer learning has been applied to music classification tasks as a pre-training step. Also in \cite{gideon2017progressive}, Gideon et al. demonstrate that emotion recognition tasks can benefit from advanced representations learned from paralinguistic tasks. This signals that emotional representation and paralinguistic representation are related to some degree. Thus it occurs to us that learning representations from emotional recognition tasks might as well benefit the escalation detection task. Recent researches on discrete class emotion recognition mainly focus on  emotions including happiness, anger, sadness, frustration and neutral \cite{busso2008iemocap}.

\subsection{Textual Embeddings}
Emotions have many ways to be expressed. As shown in \cite{busso2004analysis}, multimodal determination has become increasingly important in emotion recognition. In the ESS Dataset \cite{Schuller21-TI2}, manual transcriptions for the conversations are also provided besides the audio signals. In \cite{reimers2020making,reimers2019sentence}, Reimers et al. proposed Sentence-BERT (SBERT) to extract sentimentally meaningful sentence embeddings. Using conversation scripts as input, we encoded the textual conversations into fixed-length representations, utilizing the pre-trained multilingual model.

\section{Datasets and Methods}

\begin{figure*}
  \center
  \includegraphics[width=130mm]{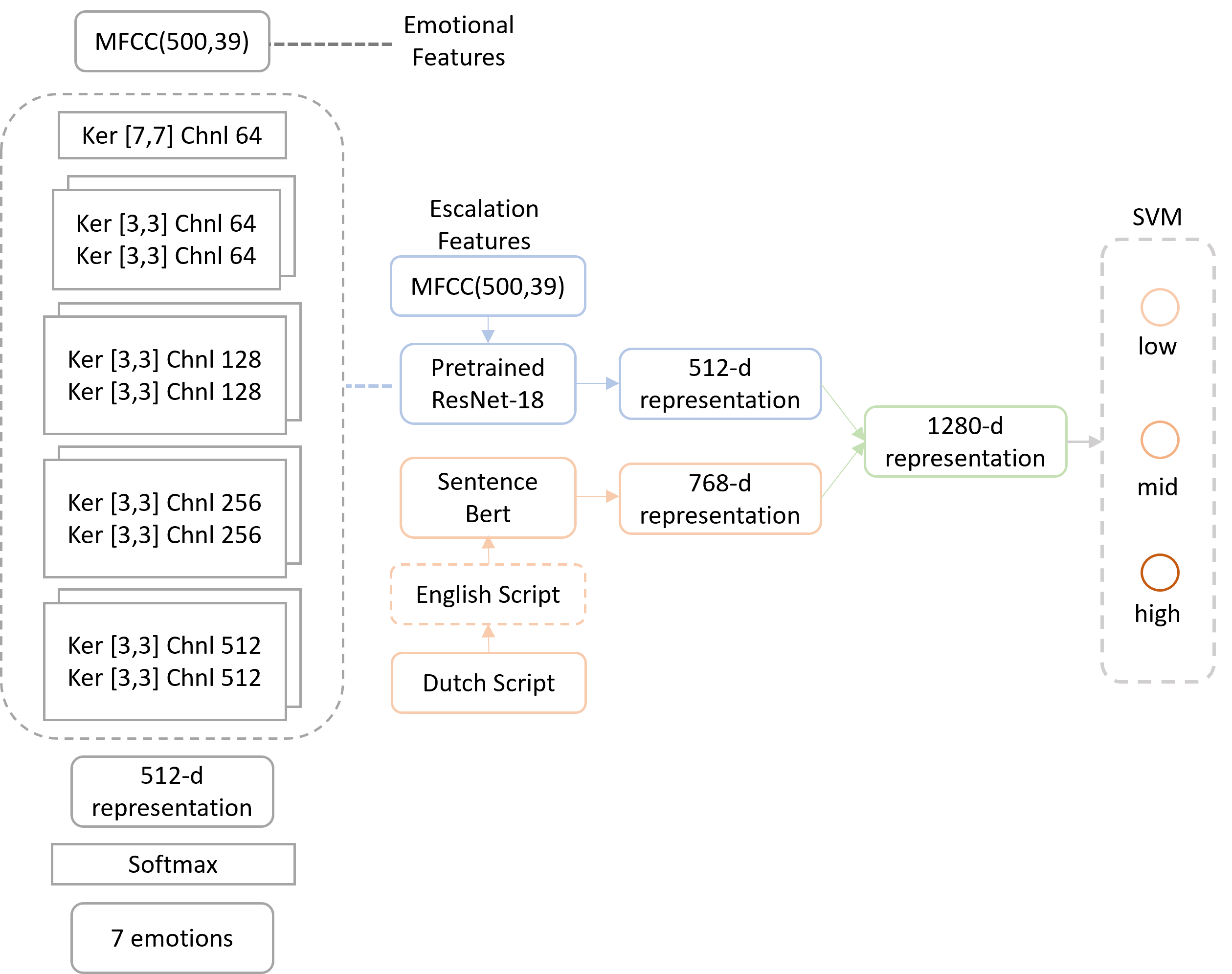}
  \caption{Pipeline of the Escalation Detection System}
  \label{fig:pipline}
\end{figure*}

An overview of our solution is shown in Figure\ref{fig:pipline}. First, we apply librosa toolkit to extract Mel Frequency Cepstral Coefficient (MFCC), which are then feed to the pre-trained ResNet-18 \cite{he2016deep} model to learn representations. The embedding extractor is pre-trained on several emotional datasets, learning emotional features from a larger source.

\subsection{Datasets}
\subsubsection{ComParE ESS Dataset}
For the ESS task, the escalation corpus consists of Dataset of Aggression in Trains (TR) \cite{lefter2013comparative} and Stress at Service Desk Dataset (SD) \cite{lefter2014audio}. The TR dataset monitors the misbehaviors in trains and train stations, and the SD dataset consists of problematic conversations that happened at the service desk \cite{Schuller21-TI2}. The escalation level has been classified into three levels, low, mid, and high. Moreover, the Dutch datasets have an average of 5 seconds for each clip. The SD dataset is used for training, and the TR dataset is used for testing. More details regarding these datasets can be found in the challenge overview paper \cite{Schuller21-TI2}.

\subsubsection{Emotional Audio Datasets}
According to previous work \cite{gideon2017progressive}, we assume that the escalatioin level detection task and semantic recognition tasks share certain patterns in acoustic representations.  Thus, we selected four well-known audio emotional datasets for joint sentimental analysis: \textbf{RAVDESS} \cite{livingstone2018ryerson} is a gender-balanced multimodal dataset. 7356 pieces of audios are carefully and repetitively labeled, containing emotions like calm, happy, sad, angry, fearful, surprise, and disgust. \textbf{CREMA-D} \cite{cao2014crema} is a high quality visual-vocal dataset, containing 7442 clips from 91 actors. \textbf{SAVEE} \cite{fayek2015towards} is a male-only audio dataset with same emotional categories with RAVDESS Dataset. \textbf{TESS} \cite{dupuis2010toronto} is a female-only audio dataset which is collected from two actresses whose mother languages are both English. TESS contains 2800 audios covering same emotional categories mentioned above.

With four audio emotional datasets combined together, we have 2167 samples for each of Angry, Happy and Sad, 1795 samples for Neutral, 2047 samples for Fearful, 1863 samples for Disgusted and 593 samples for Surprised emotions.

\subsection{Methods}
\subsubsection{Voice Activity Detection}
Voice activity detection is known as a process of identifying the presence of human speech in an audio clip \cite{ko2018limiting}. The SD dataset \cite{lefter2014audio} is collected at a service desk and therefore contains background noises. In case that the background noise undermines the paralinguistic representations, we implement the WebRTC-VAD \cite{webrtc_2017} tool to remove non-speech segments from the audio before feeding the pieces with human voice to the feature extractor.
\subsubsection{Transfer Learning}
Transfer learning has proved effective in boosting performance on low-resource classification tasks. \cite{zhao2017research}. Under our previous assumption, emotional features are important indicators in escalation level detection, hence we expect our model to include the capatility to involve emotional patterns into the process of representation learning. The emotional datasets mentioned in 3.1.2 are combined to train the ResNet-18 on the emotional recognition model. The emotional classifier and the escalation level classifier share the same configuration. After that, the pre-trained model is fine-tuned on the escalation dataset.
\subsubsection{Features}
Automatic emotional and escalation detection has been a challenging task for the fact that emotion can be expressed in multiple modalities \cite{8070966}. In multimodal emotional recognition, visual, audio and textual features are the most commonly studied channels. In our task, raw Dutch audios and manually scripted texts are provided. Thus acoustic and lexical features can be merged to jointly determine the escalation predictions. To extract the acoustic features, we first apply the WebRTC-VAD \cite{webrtc_2017} to filter out the low-energy segments. Next, MFCCs are calculated from the filtered fragments. After that, the MFCCs are passed into the emotional recognizer, which is previously jointly pre-trained on the four emotional datasets. Inside the model architecture, we also adopted the structure of the Global Average Pooling (GAP) layer \cite{lin2013network}, granting us the flexibility to accept variant length of input during evaluation phase. Our work does not construct an end-to-end detection system, instead, we employ Support Vector Machine (SVM) \cite{lin2013network} to conduct the backend classification task. According to Tang \cite{tang2015deep}, simply by replacing fully connected layers with linear SVMs can improve classification performance on multiple image classification tasks.

For the textual embeddings extraction, we adopt the pre-trained multilingual Universal Sentence Encoder (USE) \cite{yang2019multilingual} from Sentence-BERT(SBERT) \cite{reimers2020making,reimers2019sentence} to extract the sentence-level embeddings. We also compared the Unweighted Average Recall (UAR) metric by extracting Dutch embeddings directly and by extract embeddings from Dutch-to-English translation. The result shows that the embeddings from English translation outperformed the original Dutch embeddings, thus we adopted the former as textual embeddings.

\section{Experimental Results}
For the Escalation Sub-challenge, we aim to build a multimodal model to determine whether the escalation level in given conversation is low, medium or high. The official evaluation metric for ESS task is UAR \cite{Schuller21-TI2}. In this section, we will introduce our experiment setup, results and several implementation details.

\subsection{Setup}
In the audio preprocessing stage, we first applied open source tool WebRTC-VAD \cite{webrtc_2017} to filter out the unvoiced segments in the audios from the temporal domain. The noise reduction mode of WebRTC-VAD is set to 2. Next, we extract MFCCs from the filtered audios. The window length of each frame is set to 0.025s, the window step is initialized to 0.01s, and the window function is hamming function. Number of mel filters is set to 256. Also, the frequency range is from 50Hz to 8,000 Hz. The pre-emphasize parameter is set to 0.97. The representation dimension is set to 512. As for the emotional classification task, both the architecture and the configuration of the representation extractor are the same with the escalation model, except that the former is followed by a fully connected layer that maps an 128 dimension representation to 7 dimension probabilities vector produced by softmax layer, and the latter is followed by a linear SVM classifer of three levels. Weighted Cross-Entropy Loss is set as the loss function due to data imbalance. The optimizer is Stochastic Gradient Descent (SGD), with the learning rate set to 0.001, weight decay set to 1e-4, and momentum set to 0.8. The maximum training epochs is 50 epochs, with an early stop of 5 non-improving epochs. In the fine-tuning stage, the system configurations remain unchanged, except that the training epochs is extended to 300 epochs and no momentum is applied to the optimizer to reduce overfitting.

The dimension of textual embeddings extracted from the S-Bert is 768-d \cite{reimers2020making}. In the fusion part of our experiment, textual embeddings will be concatenated with the audio representations, forming a 1280-dimension of embeddings for each utterance.

\subsection{Results}
Before we train our escalation detection model on the ESS dataset, we first train the network on emotional datasets to learn emotional representations in audio. The highest UAR achieved by our model is 65.01\%. The model is selected as the pre-trained model to fine-tune on the Escalation dataset.  

We explored three factors that may have an impact on our final results. We start by analyzing whether voice activity detection will leverage the performance on the development set. Then, we fine-tune the ResNet-18 pre-trained on the four audio emotional datasets to analyze any notable improvement. Finally, we examine whether the fusion between textual embeddings extracted from SBERT \cite{reimers2020making,reimers2019sentence} and acoustic embeddings can bring improvement to the final results.

To evaluate the effect of VAD on the prediction result, we did several controlled experiments on the devel set. Table~\ref{tab:VAD} demonstrates the effect of VAD on various metrics. First, we calculated features from unprocessed audios and feed them into the embedding extractor. With acoustic embeddings alone, we scored 0.675 on the UAR metric using the Support Vector Machines (SVM) as the backend classifier. With all conditions and procedures remain the same, we added VAD to the audio pre-processing stage, filtering out non-speech voice segments. The result on the UAR metric has increased from 0.675 to 0.710. This shows that VAD is important to the escalation representation learning.

We believe that the escalation detection tasks, to some degree, share certain advanced representations with emotional recognition tasks. \cite{gideon2017progressive} Thus, we also experimented on fine-tuning parameters on the escalation dataset with the model pre-trained on the emotional datasets. Table~\ref{tab:TL} shows the experiment results after implementing transfer learning to our system. We have witnessed a positive impact of VAD on our experiment results on the devel set, thus the base experiment has been implemented with VAD applied. We can see that, after applying the pre-trained model to the MFCC+VAD system, with acoustic embeddings alone, the score reached 0.810 on the metric UAR, which turns out to be a significant improvement. This also proves that the emotional features can benefit paralinguistic tasks by transfer learning. Additionally, the MFCC+VAD+PR system may have already been stable enough that there is no noticeable improvement brought by the textual embeddings fusion. 

Other than the experiments recorded above, we also implemented a wide range of trials involving different features, networks, and techniques. As listed in Table~\ref{tab:OtherExperiments}, we recorded other meaningful experiments with convincing performance on the devel set that could be applied to final model fusion. Other common acoustic features like the Log filterbank are also under experiment. Label Smoothing technique \cite{muller2019does} is applied on the MFCC+VAD+PR model but brought a slightly negative impact. According to the experiment results, the Voice Activity Detection has again been proved effective in enhancing model performance on the devel set. ResNet-9 without being pre-trained is also implemented, the classification result is 74.9\% UAR on the devel set.

To further improve our model performance, we proposed model fusion on three models of the best performance on the devel set. The fusion is conducted in two ways, early fusion, and late fusion. In the early fusion stage, we also proposed two approaches to deal with the embeddings. The first approach is concatenating the embeddings and the second is simply taking the mean value of the embeddings, both scenarios are followed by the SVM as the classifier. As for the late fusion, we proposed a voting mechanism among the three models' decisions. The fusion result is shown in Table~\ref{tab:modelfuse}. By implementing late fusion, we managed to get the best system performed on the devel set with UAR score 81.5\%.

\begin{table}[t]
  \caption{Effects of Voice Activity Detection (VAD) on the devel set. \textbf{TE}: Textual Embeddings fused.}
  \label{tab:VAD}
  \centering
  \begin{tabular}{ lclclcl }
    \toprule
    Model Name & Precision & UAR    & F1-Score      \\
    \midrule
    MFCC                & 0.640   & 0.675   & 0.647   \\
    MFCC+VAD            & 0.675   & 0.710   & 0.688   \\
    MFCC+TE             & 0.652   & 0.690   & 0.664   \\
    MFCC+VAD+TE         & 0.676   & 0.721   & \textbf{0.691}   \\
    Baseline Fusion \cite{Schuller21-TI2}     & -       & \textbf{0.722}   & - \\
    \bottomrule
  \end{tabular}
\end{table}

\begin{table}[t]
  \caption{Effects of fine-tune of pre-trained ResNet-18 on devel set. \textbf{PR}: Pre-trained ResNet-18 applied. }
  \label{tab:TL}
  \centering
  \begin{tabular}{ lclclcl }
    \toprule
    Model Name & Precision & UAR    & F1-Score      \\
    \midrule
    MFCC+VAD            & 0.675   & 0.710   & 0.688   \\
    MFCC+VAD+PR         & 0.807   & \textbf{0.810}   & 0.788   \\
    MFCC+VAD+PR+TE  & 0.807   & \textbf{0.810}   & 0.788   \\
    Baseline Fusion \cite{Schuller21-TI2}    & -       & 0.722   & - \\
    \bottomrule
  \end{tabular}
\end{table}

\begin{table}[t]
  \caption{Extra Experiments. \textbf{LS}: Label Smoothing. }
  \label{tab:OtherExperiments}
  \centering
  \begin{tabular}{ lclclcl }
    \toprule
    Model Name & Precision & UAR    & F1-Score      \\
    \midrule
    Logfbank            & 0.670   & 0.743   & 0.684   \\
    Logfbank+VAD        & 0.711   & 0.778   & 0.733 \\
    MFCC+VAD+PR+LS      & 0.781   & \textbf{0.781}   & \textbf{0.761} \\
    MFCC+VAD+ResNet-9   & 0.727   & 0.749   & 0.725   \\
    Baseline Fusion \cite{Schuller21-TI2}    & -       & 0.722   & - \\
    \bottomrule
  \end{tabular}
\end{table}

\begin{table}[t]
  \caption{Model Fusion. Three models: MFCC+VAD+PR, MFCC+VAD+PR+LS, Logfbank+VAD }
  \label{tab:modelfuse}
  \centering
  \begin{tabular}{ lclclcl }
    \toprule
    Fusion Approach & Precision & UAR    & F1-Score      \\
    \midrule
    Concatenate      & 0.783    & 0.800    & 0.779   \\
    Mean             & 0.789    & 0.805    & 0.789 \\
    Voting      & 0.810   & \textbf{0.815}   & \textbf{0.803} \\
    \bottomrule
  \end{tabular}
\end{table}

\section{Discussion}

Our proposed best fusion model exceeded the devel set baseline by 12.8\%. It is worth mentioning that the WebRTC-VAD system is not able to tease out every non-speech segment. Instead, its value cast more light on removing the blank or noisy segments at the beginning and end of an audio clip. This is reasonable since the unsounded segments in a conversation are also meaningful information to determine the escalation level and emotion. Dialogues would be more likely labeled as high escalation level if the speaker is rushing through the conversation and vice versa. Thus we agreed that a more complicated Neural-Network-based voice activity detector may be unnecessary in this task.

The significant improvement of our system on the devel set has again proved that emotional recognition features and paralinguistic features share certain advanced representations. Just as we mentioned in the related work part, they benefit from each other in the transfer learning tasks. However, due to the small scale of the training set, the overfitting problem is highly concerned. Thus we chose ResNet-18 to train the model on a combined emotional dataset, containing 12,000+ labeled emotional clips. This architecture has also been proved to be effective in the Escalation detection task.

For this task, we adopted Sentence-BERT \cite{reimers2019sentence} as the textual embeddings extractor. We utilized the pre-trained multilingual BERT model which is capable of handling Dutch, German, English, etc. We chose to translate the raw Dutch text to English text before feeding them into the embedding extractor, for we did a comparative experiment, which showed that the English textual embeddings alone significantly outperformed the Dutch textual embeddings. The UAR on the devel set achieved by English Embeddings alone is around 45\%, whose detection ability is very likely to be limited by the dataset scale and occasional errors in translation. Had we have richer textual data, the lexical embeddings should be of more help.

An unsuccessful attempt is adding denoising into the preprocessing attempt. Denoising should be part of the preprocessing stage since most of the collected audios contain background noise from public areas. According to \cite{letcher2018automatic}, they first denoise the police body-worn audio before feature extraction, which turns out rewarding for them in detecting conflicts from the audios. However, our attempt does not improve the performance. Our denoised audio is agreed to be clearer in human perception and contains weaker background noises. However, the performance on the devel set is significantly degraded. Our assumption is that, unlike the police body-warn audio, which mostly contains criminality-related scenarios, the conversation audios in ESS dataset happen with richer contextual environments. The speech enhancement system might also affect the signal of speech which might be the reason for the performance degradation.

\section{Conclusions}

In this paper, we proposed our solution to the ComParE 2021 Escalation Detection Challenge \cite{Schuller21-TI2}. We applied Voice Activity Detection to pre-process the ESS Dataset. We also pre-trained a ResNet-18 architecture on an emotional recognition task and fine-tune the parameters with the escalation dataset. We also validated that learning emotional representations can leverage the ability of the model to detect conversational escalation situations. By involving textual information in the classification process, the model can become more stable and robust. The single best model can achieve 81.0\% on the devel set. By doing the late fusion the models after fusion are able to achieve the 81.5\% UAR. Future efforts will be focusing on addressing the over-fitting problem.

\section{Acknowledgements}

This research is funded in part by the National Natural
Science Foundation of China (61773413), Key Research
and Development Program of Jiangsu Province
(BE2019054), Six talent peaks project in Jiangsu Province (JY-
074), Science and Technology Program of Guangzhou City
(201903010040,202007030011).

\bibliographystyle{IEEEtran}
 
\bibliography{mybib} 


\end{document}